# Analysis of Light-Weight Cryptography Algorithms for UAV-Networks

Aanchal Patel  Aswani Kumar Cherukuri

School of Computer Science Engineering and Information Systems,

Vellore Institute of Technology, Vellore 632014, India. Email: cherukuri@acm.org

Abstract

Unmanned Aerial Vehicles (UAVs) are increasingly utilized across various domains, necessitating robust security measures for their communication networks. The ASCON family, a NIST finalist in lightweight cryptography standards, is known for its simplistic yet resilient design, making it well-suited for resource-constrained environments characterized by limited processing capabilities and energy reservoirs.  This study focuses on understanding the integration and assessment of the ASCON encryption algorithm in UAV networks, emphasizing its potential as a lightweight and efficient cryptographic solution.  The research objectives aim to evaluate ASCON variants' effectiveness in providing security comparable to AES-128 while exhibiting lower computational cost and energy consumption within simulated UAV network environments. Comparative analysis is conducted to assess performance metrics such as encryption and decryption speeds, resource utilization, and resistance to cryptographic vulnerabilities against established algorithms like AES.  The proposed architecture outlines the setup for evaluating ASCON and AES in a simulated UAV network environment, focusing on performance monitoring, security analysis, and comparative studies. Performance metrics including peak and average execution times, overall throughput, and security properties against various cryptographic attacks are measured and analysed to determine the most suitable cryptographic algorithm for UAV communication systems.  Performance results indicate that ASCON-128a as the optimal choice for UAV communication systems requiring a balance between efficiency and security. Its superior performance metrics, robust security properties, and suitability for resource-constrained environments position it as the preferred solution for securing UAV communication networks. By leveraging the strengths of ASCON-128a, UAV communication systems can achieve optimal performance and security, ensuring reliable and secure communication in challenging operational environments.

## 1. INTRODUCTION

Traditional cryptographic algorithms, like AES-128 [2], impose a significant computational and energy burden on resource-constrained UAV networks. This research aims to address the challenge of balancing robust communication security with efficient resource utilisation within UAV networks. Specifically, we investigate the effectiveness of the ASCON family, a lightweight cryptography algorithm recently standardised by NIST [3], in providing comparable security with respect to AES-128 while significantly reducing computational overhead and energy consumption.

By evaluating the performance of various ASCON variants and AES in a simulated UAV network environment, this research seeks to determine if any variant from the ASCON family offers a viable alternative to AES for comparably secure and efficient UAV communication. The rapid proliferation of unmanned aerial vehicles (UAVs), or drones, across various sectors underscores the critical need for efficient and secure communication networks [Fig.1] to support their operations. UAVs are increasingly deployed in diverse applications, ranging from military reconnaissance and surveillance to precision agriculture, infrastructure inspection, and delivery services. In these contexts, UAVs serve as essential tools for collecting data, monitoring remote locations, and performing tasks that are otherwise difficult or hazardous for humans to undertake.

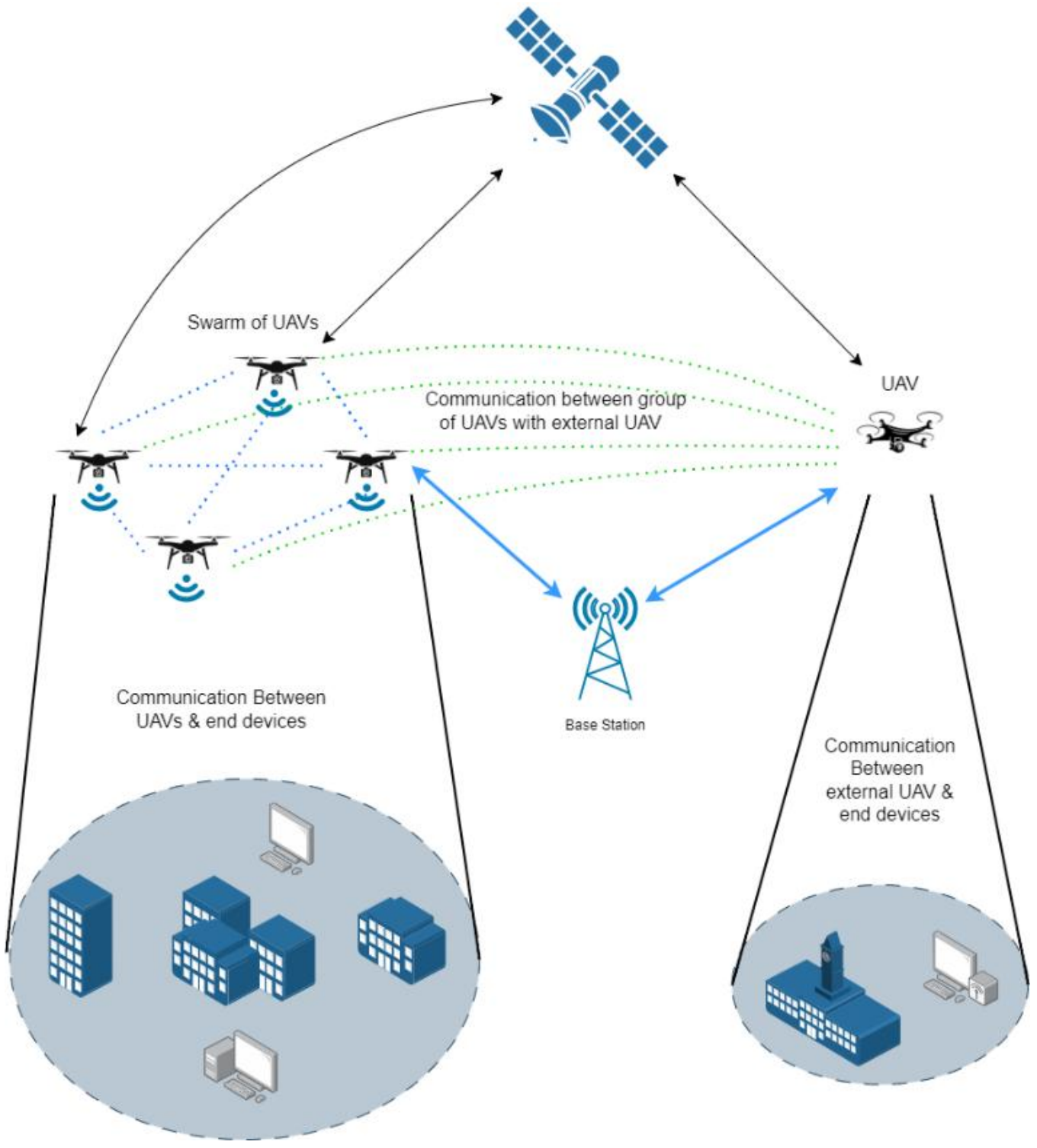

*Fig.1: Data transmission in secured UAV network*

Central to the effectiveness and safety of UAV operations is the establishment of robust communication links between drones and ground control stations. These communication networks enable real-time command and control, data transmission, and telemetry exchange, allowing operators to remotely pilot UAVs and receive sensor data for analysis. However, the reliance on wireless communication exposes UAV networks to various security threats, including eavesdropping, data interception, signal jamming, and cyberattacks.

Given the sensitive nature of the data transmitted within UAV networks, ensuring the confidentiality, integrity, and authenticity of communications is paramount. Any compromise in the security of these networks could have severe consequences, ranging from unauthorized access to sensitive information to the manipulation of UAV behaviour or even the hijacking of drone operations for malicious purposes. Moreover, as UAV technology continues to advance and drones become increasingly integrated into the broader Internet of Things (IoT) ecosystem, the complexity and scale of UAV communication networks are expected to grow significantly. This expansion further amplifies the importance of addressing security challenges proactively to mitigate potential risks and vulnerabilities inherent in UAV deployments.

Therefore, the motivation behind this research lies in the urgent need to develop and evaluate cryptographic solutions tailored to the unique requirements of UAV communication networks. By leveraging lightweight encryption algorithms and advanced security protocols, we aim to enhance the resilience of UAV networks against emerging threats while minimizing the computational overhead and resource constraints typical of drone platforms.

Unmanned aerial vehicles (UAVs), commonly referred to as drones, have emerged as indispensable tools across a wide range of industries. In agriculture, drones are used for crop monitoring, irrigation management, and precision agriculture, enabling farmers to optimize yields and reduce resource usage. Infrastructure inspection tasks, such as monitoring pipelines, bridges, and power lines, benefit from the aerial perspective provided by UAVs, which can access difficult-to-reach locations with ease. In the realm of surveillance and security, UAVs play a crucial role in border patrol, disaster response, and law enforcement operations, providing real-time situational awareness and reconnaissance capabilities. The proliferation of UAVs has led to their integration into the broader Internet of Things (IoT) ecosystem, where they interact with sensors, actuators, and other IoT devices to collect and exchange data. This convergence of UAV technology with IoT systems presents new opportunities for data-driven decision-making and automation but also introduces complex security challenges.

One of the primary concerns in UAV communication networks is the vulnerability of data transmissions to interception and tampering [Fig.2]. As UAVs transmit sensitive information, such as video feeds, sensor data, and mission-critical commands, securing these communication channels becomes very important. Traditional encryption standards, such as the Advanced Encryption Standard (AES), offer robust security but may impose significant computational overhead on resource-constrained UAV platforms. To address the limitations of traditional encryption algorithms in UAV environments, lightweight cryptography has emerged as a promising solution. Lightweight encryption algorithms are specifically designed to operate efficiently on devices with limited processing power and memory, making them well-suited for UAVs and other IoT devices. These algorithms prioritize efficiency while still providing strong cryptographic protection, making them ideal candidates for securing UAV communication networks.

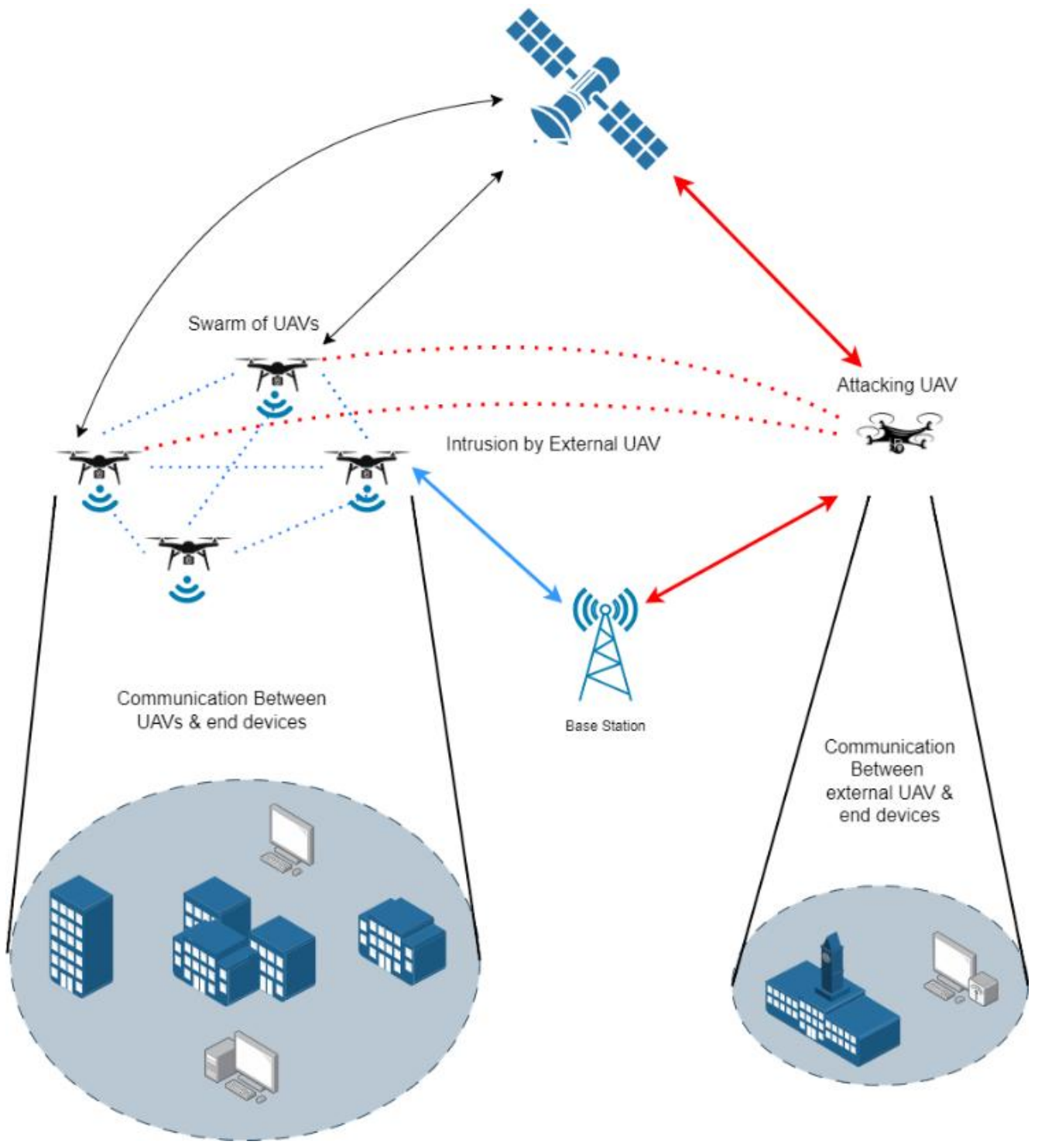

*Fig.2: Intrusion Within UAV Network by External UAV*

By delving into the realm of lightweight cryptography and evaluating its suitability for UAV applications, this research seeks to bridge the gap between security requirements and resource constraints in UAV communication networks. Through a comprehensive analysis of cryptographic methods and their performance in UAV scenarios, this study aims to contribute to the development of secure and efficient communication protocols for the next generation of UAV systems.

## 2. LITERATURE SURVEY

This section addresses an in-depth analysis of existing approaches relevant to our research, highlighting their limitations.

Gupta Jain et al. [4] conduct an extensive overview of the opportunities and challenges associated with UAV communication networks. They emphasise the growing importance of UAV communication networks in civil applications and highlighting the necessity of robust and integrated systems to effectively facilitate UAV operations by emphasizing on dynamic topology, intermittent links, and power and bandwidth constraints.

Three key areas are the focus of the survey: energy efficiency, seamless handover, and routing. Gupta et al. delve into the limitations of various approaches, including static, proactive, reactive, hybrid, and delay-tolerant protocols, for tackling the dynamic nature of UAV networks. Additionally, it highlights the lack of research on seamless handover mechanisms catered specifically for UAV environments, which poses a significant challenge to preserving uninterrupted communication during network transitions. Furthermore, the researchers illustrate the importance of energy-efficient protocols in maximizing the operational lifespan of UAV networks, recommending strategies such as transmission power control and load distribution to optimize resource consumption.

While the study provides invaluable insight into the current state of research in UAV communication networks, it is essential to recognise some limitations. First off, as UAV technology is continually evolving, the survey may not cover all current developments and breakthroughs in the industry. Furthermore, the discussion on routing protocols could derive value from an in-depth study of the specific obstacles and trade-offs related to each individual protocol. likewise, the survey could explore further into potential solutions or new directions of study to address the shortcomings and challenges that have been found. Regardless these drawbacks, Gupta et al. provide an in-depth review of the significant issues with UAV communication networks, paving the stage for subsequent research in this emerging area.

Existing encryption methods like ChaCha20-Poly1305 are insufficient in such environments, risking real-time performance degradation making packet lose common. To set off these limitations T. Li et al. [5] proposes a lightweight secure communication scheme for UAV networks, addressing challenges like packet loss in dynamic battlefield conditions by introducing an innovative approach based on the Counter mode of the SM4 algorithm, known for its resilience to packet loss. The scheme includes key update mechanisms, encryption/decryption processes, and data integrity checks. For security purposes, session keys are exchanged and updated on a regular basis, and the integrity of messages is guaranteed via HMAC. Its effectiveness is confirmed by evaluation in a UAV network environment, where it outperforms conventional encryption techniques. Designed with resource-constrained UAV networks in mind, the method ensures data integrity and confidentiality in dynamic and potentially hostile situations by supporting real-time communication and tolerating packet loss.

Kamil, Muhammed et al [6] presents an innovative certificate-less integration technique optimised for resource-constrained drones that uses elliptic curve cryptography (ECC) for key formation. ECC's advantages in terms of security, efficiency, and communication are highlighted via comparative study. In addition to providing safe session key exchange and defence against cyberattacks, the system architecture guarantees memory efficiency and quick arithmetic operations.

AES's vital role in guaranteeing data security across a range of applications including IOT and drones is highlighted by Abdullah et al. [7], who also note the algorithm's effectiveness and efficiency in comparison to other encryption techniques. domains of AES implementation in both software and hardware settings, including its integration into cryptographic protocols like SSL and TLS. As a result, Cecchinato et al. [8] used lightweight AES encryption techniques to secure sensitive audio and video data carried by UAVs. Comprehensive testing confirms that encryption and decryption perform properly even in difficult situations. Because AES cryptography is integrated at the application level, the technology is suitable for resource-intensive UAV systems. The results show that the encryption is effective in real-time even under challenging transmission settings. The technique encrypts data at each packet to guarantee decryption integrity. Even in a different work, Cecchinato et al. [9] details the data collection process using drones equipped with cameras and microphones, as well as the ground station processing. Secure communication is used to protect the confidentiality and integrity of data. Because drones have low computational capacity, lightweight AES encryption is used in conjunction with MAC encryption for wireless communication. The experiment demonstrates that encryption and decryption protocols may be improved to constantly protect data, even under adverse conditions.

Even though AES (Advanced Encryption Standard) is widely used in UAV networks because of all these qualities, Zhong et al. [10] found that its main flaw is that it is susceptible to side-channel attacks, particularly those that depend on timing and power usage analysis. Rather than target the algorithm directly, these attacks take use of the physical aspects of the cryptographic implementation. Attackers may be able to determine the secret key being used by examining how long or how much power AES operations consume, which could jeopardise the encryption's security.

These leads are used in the search for a novel cryptographic method appropriate for the resource-constrained UAV environment. A competition was launched in 2019 by NIST [14] [11] [12] [13] to standardise a new lightweight cryptography algorithm. Various software applications are being considered for ASCON, GIFT-COFB, ISAP, PHOTON-Beetle, Romulus, SPARKLE, TinyJAMBU, and Xoodyak, which have shown performance benefits over NIST standards in software benchmarks. The top contenders ASCON, Elephant, GIFT-COFB, PHOTON-Beetle, Romulus, TinyJAMBU, and Xoodyak showed performance improvements above NIST standards for hardware applications. Promising features are offered by ISAP for applications that need side-channel resistance.

Mohajerani et al. [15] assessed for Lightweight Cryptography (LWC) candidates' side-channel resistance, a critical need in the NIST Standardisation procedure. When there are fewer finalists, it is crucial to evaluate how this feature will affect real-world applications. The platform incorporates distributed creation of protected hardware and software implementations, dynamic matching of evaluators and implementers, and self-identification and characterization of evaluation labs. Xilinx Artix-7 FPGAs are used to benchmark hardware implementations [17], creating classes with comparable side-channel resistance. Because of their distinct benefits in terms of throughput, area, throughput-to-area ratio, or randomization requirements, four contenders are identified: Ascon, Xoodyak, TinyJAMBU, and ISAP. The environment incorporates distributed development of protected hardware and software implementations, dynamic matching of evaluators and implementers, and self-identification and characterization of evaluation labs. Xilinx Artix-7 FPGAs are used to benchmark hardware implementations, creating classes with comparable side-channel resistance. Because of their distinct benefits in terms of throughput, area, throughput-to-area ratio, or randomization requirements, four contenders are identified: Ascon, Xoodyak, TinyJAMBU, and ISAP. According to Elsadek et al. [16] TinyJambu for short burst message applications and Xoodyak for continuous long message processing, underscoring the importance of tailored hardware for LWC algorithms in enabling secure and efficient communication on resource-constrained devices.

To meet the security requirements of devices with limited resources, particularly those operating on the Internet of Things (IoT) space, the National Institute of Standards and Technology (NIST) [18] standardised the Ascon family of lightweight cryptographic algorithms. Ascon's adaptability includes variable-length hash functions and shorter integrity tags, allowing developers to customize security levels efficiently. Its set of features, which includes support for signature schemes like Ed25519 and ECDSA and plans for API integration, makes cryptographic implementations on limited devices easier. It covers AEADs, hash functions, KDFs, PBKDFs, and MACs. By eliminating weaker key possibilities, Ascon improves post-quantum security and establishes itself as a flexible solution for successfully preserving devices with limited resources. New PRF and MAC implementations are brought to Ascon through enhancements by Dobraunig et al. [19], which promise lightweight efficiency and strong security against key recovery efforts. M. Tanveer et al. [20] used PASKE-IoD with ASCON, which offers strong security features designed for IoD architecture as well as anonymity in IoT and Internet of Drones (IoD) environments.

Even Chakraborty et al. [21] considers ASCON as desirable choice for lightweight cryptography applications due to its unique properties, which include double-keyed initialization and finalisation procedures that enhance its security and effectiveness. Furthermore, Kamil et al. [22] suggested it as a crucial component of lightweight encryption APIs for Internet of Things applications, guaranteeing safe communication while skilfully managing resource limitations. According to Sliwa et al. [23] [24] Underwater communication scenarios benefit from Ascon's adaptability for resource-constrained environments due to its efficiency and security, which make it an appropriate choice for tackling the unique challenges associated with secure communication in such environment.

To reduce security concerns like jamming and eavesdropping, V. K. Ralegankar et al. [25] underline the urgent need for secure communication solutions in UAV networks and offer a novel architectural approach that makes use of quantum cryptography. Sharma and Mehrasuch et al. [26] recommend implementing post-quantum cryptography and lightweight encryption, as well as integrating advanced network types like 5G and IoT, malware identification, and intrusion detection systems. Mekdad, Yassine, et al. [27] highlight privacy issues related to drones that have cameras installed, arguing in favour of all-encompassing security solutions, safe coding techniques, and strong defences against new threats. Physical-Layer Security (PLS) approaches are recommended by Xu, Fang, et al. [28] to improve privacy in UAV communications. SecAuthUAV, a lightweight mutual authentication protocol, is introduced by T. Alladi et al. [29] resilient against various security attacks.

Norzailawati et al. [30] discuss on how UAVs will help to enable future mobile networks, suggesting that technologies like SDN and NFV be used to solve problems with interoperability and management. Unresolved research concerns are listed by Oubbati et al. [31] and include optical communication, wireless power transmission, security in 6G networks, and AI integration for network optimisation in UAV-assisted networks. The overall goal of the survey was to improve communication reliability and data security in dynamic and potentially hostile environments by analysing current methods, identifying their shortcomings, and putting forth creative solutions to the security and efficiency issues in UAV communication networks.

## 3. PROPOSED ARCHITECTURE

This architecture [fig.3] aims to evaluate the performance and security of all ASCON variations and AES encryption algorithms in a simulated UAV network environment. It focuses on providing a comprehensive comparative analysis considering various factors relevant to resource-constrained UAV communication.

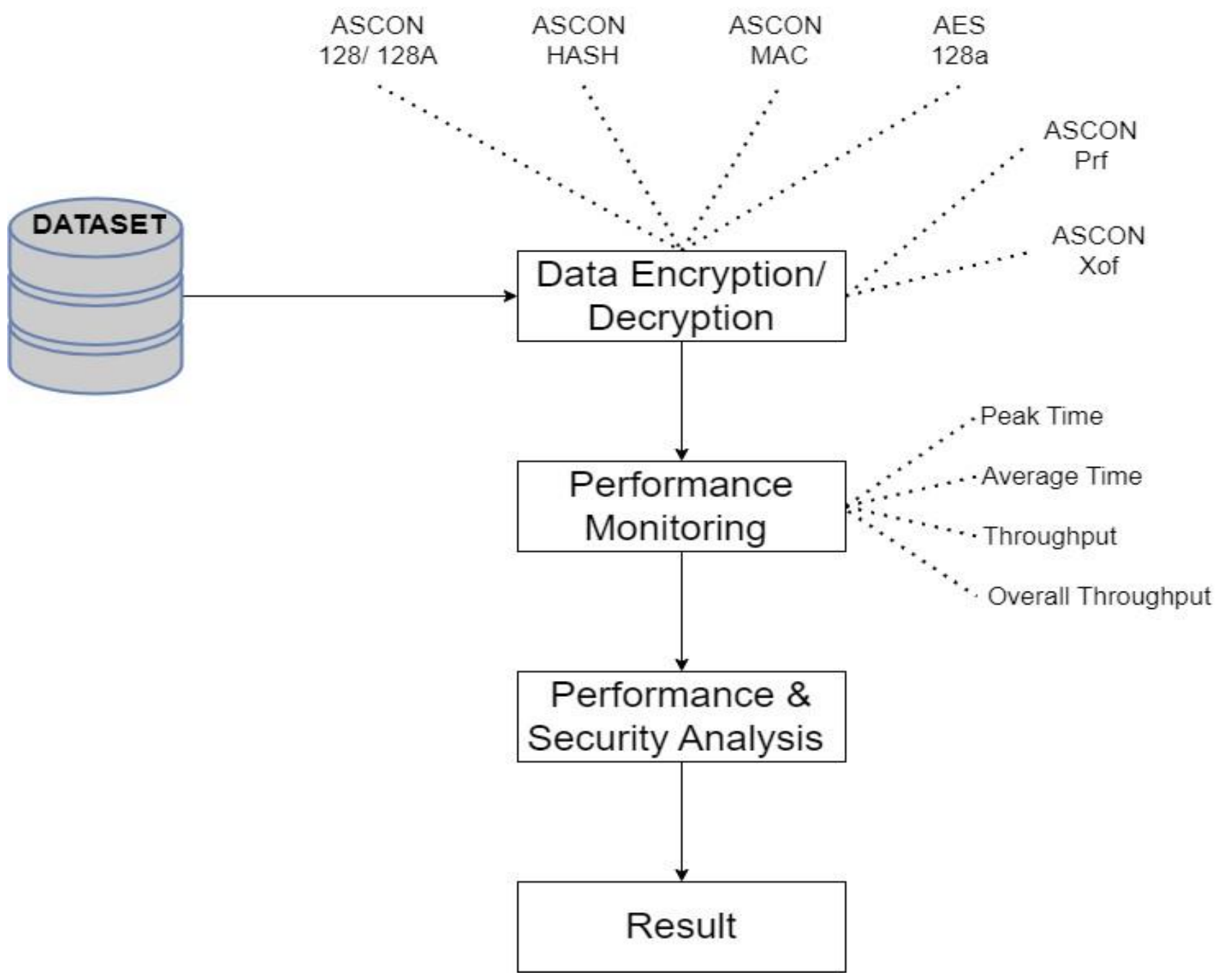

*Fig.3: Proposed Architecture*

3.1 ASCON FAMILY:

The Ascon family is a set of lightweight cryptographic algorithms designed for efficiency in constrained environments like resource-limited devices. The ASCON design is based on a sponge construction along the lines of SpongeWrap and MonkeyDuplex giving it versatility, security like collision resistance and efficiency. Various components of ASCON family are:

- **Authenticated Encryption with Associated Data (AEAD):** This is the core functionality of Ascon. It allows encrypting data (confidentiality) while ensuring its integrity through authentication. Associated data can be included in the authentication process for additional security.
- **Hash Functions (HASH) and Extensible Output Functions (XOF):** Hash functions take an arbitrary input of any length and generate a fixed-size output (hash). XOFs are like hash functions but can produce outputs of any desired length. These functions are useful for data integrity checks, generating random numbers, and key derivation.
- **Pseudo-Random Functions (PRF) and Message Authentication Codes (MAC):** PRFs are deterministic algorithms that map data to a seemingly random output value. MACs are a specific type of PRF used for message authentication, where a secret key is used to generate a tag that cryptographically binds to the message. Both PRFs and MACs are essential for ensuring data integrity and preventing message forgery. [32] [33] [34]

A. <u>ASCON'S PERMUTATION</u>

The foundation of Ascon's cryptographic operations is its permutation, which is shared by all Ascon family members. An SPN-based round transformation is applied iteratively in this permutation, either 12 times for $p^a$ or a configurable number of times (6 to 8) for $p^b$. It consists of three basic steps: insertion of round constants, nonlinear substitution layer, and linear diffusion layer which operates on a 320-bit state divided into five 64-bit words ($x_0$ to $x_4$). Specifically, each word undergoes a series of XOR operations with itself shifted to the right by specific bit positions, defining the permutation's structure. Whereas the nonlinear replacement layer applies a 5-bit S-box [Fig.4] 64 times in parallel, the linear diffusion layer [Fig.5] XORs rotated copies of each word.

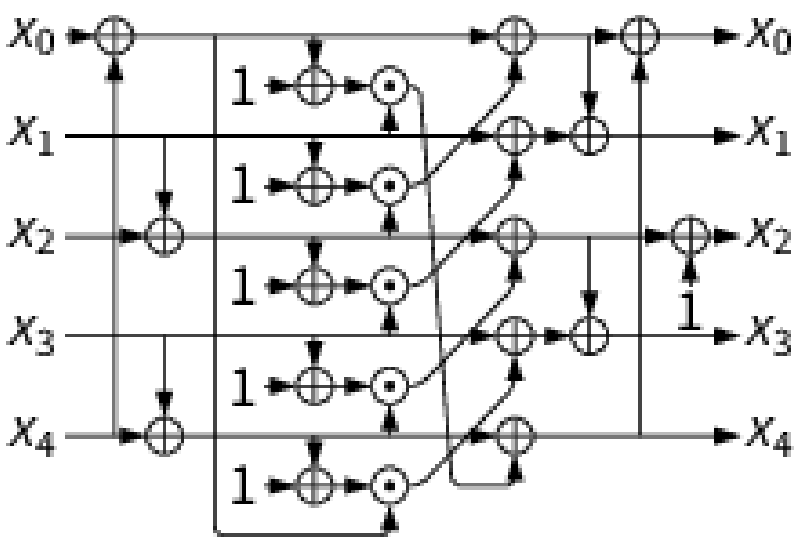

Fig.4: *Ascon's S-box*

$$X_0 = x_0 \oplus (x_0 \ggg 19) \oplus (x_0 \ggg 28)$$

$$X_1 = x_1 \oplus (x_1 \ggg 61) \oplus (x_1 \ggg 39)$$

$$X_2 = x_2 \oplus (x_2 \ggg 1) \oplus (x_2 \ggg 6)$$

$$X_3 = x_3 \oplus (x_3 \ggg 10) \oplus (x_3 \ggg 17)$$

$$X_4 = x_4 \oplus (x_4 \ggg 7) \oplus (x_4 \ggg 41)$$

*Fig.5: Ascon's linear layer*

(Ascon's permutation: $\oplus$ denotes XOR, $\odot$ denotes and, $\ggg$ is rotation to the right)

B. <u>ASCON'S AUTHENTICATED ENCRYPTION MODES</u>

Ascon uses a duplex-sponge-based approach for its authenticated encryption modes, recommending 128-bit key, tag, and nonce lengths. The sponge processes 64- or 128-bit message blocks while running in a 320-bit state. There are four stages to the encryption process: initialization, associated data processing, plaintext processing, and finalization. While associated data processing updates the state with associated data blocks, initialization seeds the state with the key and nonce. In plaintext processing, plaintext blocks are inserted into the state and ciphertext blocks are extracted; in finalization, the key is introduced once more to extract the authentication tag. During initialization and finalization, a stronger permutation ($p^a$) is applied, while the core permutation ($p^b$) is applied after every injected block except the last plaintext block. Parameters for rounds a and b, along with the sponge's rate and capacity, vary depending on the Ascon variant. The recommended parameters are in Table1.

| Cipher | Bit Size Of | | | | |
|---|---|---|---|---|---|
| | Key | nonce | tag | tag | capacity |
| Ascon-128 | 128 | 128 | 128 | 64 | 256 |
| Ascon-128a | 128 | 128 | 128 | 128 | 192 |

*Table1: Parameters for Ascon authenticated Encryption*

C. <u>ASCON'S HASHING MODES</u>

The Ascon family uses sponge-based modes [Fig.6] to provide hash functions (Ascon-Hash and Ascon-Hasha) and extensible output functions (Ascon-Xof and Ascon-Xofa) for hashing. These modes offer 128-bit security and a minimum hash size of 256 bits. Hashing processes squeeze the hash value in 64-bit blocks and absorb the message in 64-bit blocks. The complete a-round permutation $p^a$ is applied at initialization and finalisation, following the final message block, whereas the b-round permutation $p^b$ is applied to the state following each absorbed or compressed block, apart from the last.

*

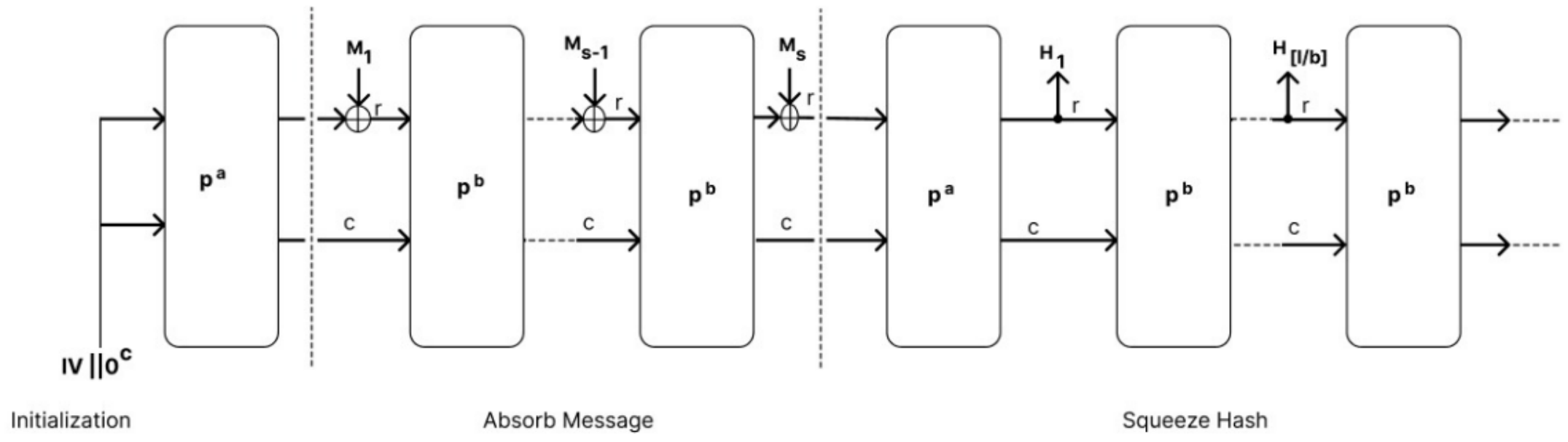

*Fig.6: The sponge mode for Ascon hashing*

The recommended parameters for hashing modes are in Table2.

| Algorithm | Bit Size Of | | | Rounds | |
|---|---|---|---|---|---|
| | hash output | rate | capacity | $p^a$ | $p^b$ |
| ASCON-HASH | 256 | 64 | 256 | 12 | 12 |
| ASCON-XOF | arbitrary | 64 | 256 | 12 | 12 |
| ASCON-HASHA | 256 | 64 | 256 | 12 | 8 |
| ASCON-XOFA | arbitrary | 64 | 256 | 12 | 8 |

*Table2: parameters for Ascon Hashing*

D. <u>ASCON ENCRYPTION AND DECRYPTION ALGORITHM</u> [35]:

| **Encryption**<br><br>$E_{k,r,a,b}$ (K, N, A, P) |
|---|
| **Input:** key K ∈ $\{0,1\}^{K}$, k < 160,<br>nonce N ∈ {0,1}128<br>associated data A ∈ $\{0,1\}^{*}$,<br>plaintext P ∈ $\{0,1\}^{*}$<br>**Output:** ciphertext C ∈ $\{0,1\}^{\|P\|}$,<br>tag T ∈ $\{0,1\}^{128}$ |
| **Initialization**<br>S ← $IV_{,r,a,b}$ \|\| K \|\| N<br>S ← $p^{a}$ (S) ⊕ ($0^{320-k}$ \|\| K)<br>**Processing Associated Data**<br>if \|A\|>0 then<br>$A_1$... As ← r-bit blocks of A\|\|1\|\|0*<br>for i= 1, ..., s do<br>S ← $p^{b}$ (($S_r$ ⊕ $A_i$) \|\| $S_c$)<br>S ← S ⊕ ($0^{319}$ \|\| 1)<br>**Processing Plaintext**<br>$P_1$ ... $P_t$ r-bit blocks of P\|\|1\|\|0*<br>for i= 1, ..., t-1 do<br>$S_r$ ← $S_r$ ⊕ $P_i$<br>$C_i$ ← $S_r$<br>S ← $p^{b}$ (S)<br>$S_r$ ← $S_r$ ⊕ $P_t$<br>$\tilde{C}_t$ ← $[S_r]_{\|P\| \bmod r}$<br>**Finalization** |

$S \leftarrow p^a (S \oplus (0^r \| K \| 0^{320-r-k})$

$T \leftarrow [S]^{128} \oplus [K]^{128}$

**return** $C_1 \| ... \| C_{t-1} \| \tilde{C}_t$ T

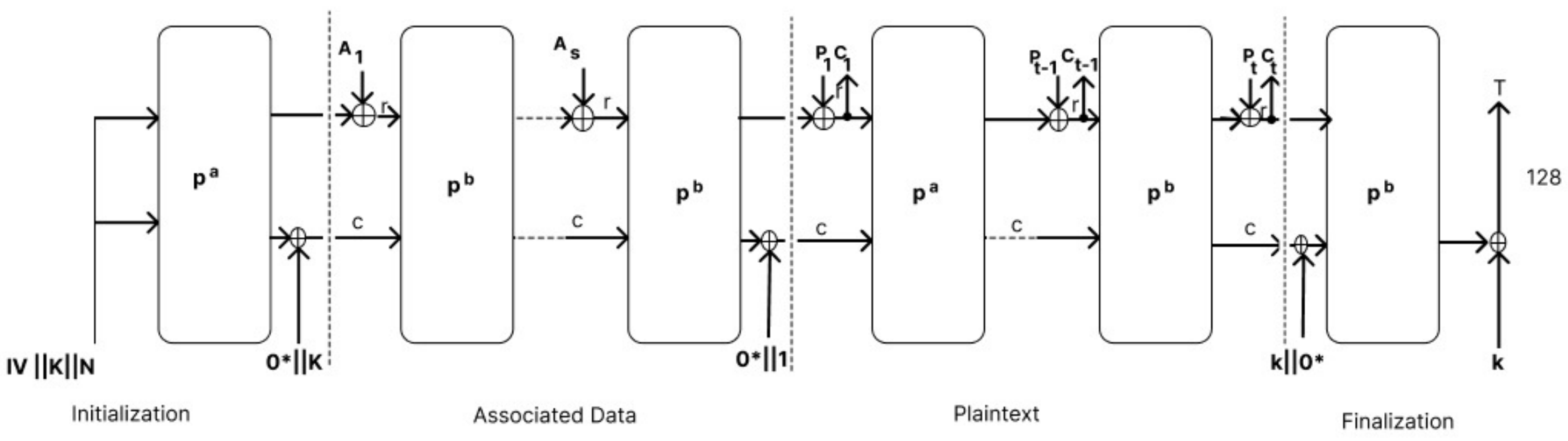

*Fig.7: Encryption in Ascon*

| **Decryption**<br><br>$D_{k,r,a,b}$ (K,N,A,C,T) |
|---|
| **Input:** key K Є $\{0,1\}^k$, k < 160,<br><br>nonce N ∈ {0,1}128,<br><br>associated data A ∈ {0,1} *,<br><br>ciphertext C ∈ {0,1} *,<br><br>tag T ∈ $\{0,1\}^{128}$<br><br>**Output:** plaintext P ∈ $\{0,1\}^{\|C\|}$ or 1 |
| **Initialization**<br><br>$S \leftarrow IV_{k,r,a,b} \| K \| N$<br><br>$S \leftarrow p^a (S) \oplus (0^{320-k} \| K)$<br><br>**Processing Associated Data**<br><br>if \|A\|>0 then<br><br>$A_1 ... A_s \leftarrow$ r-bit blocks of A\|\|1\|\|0*<br><br>for i= 1, ..., s do |

$S \leftarrow p_b\,((S_r \oplus A_i)\;||\;S_c)$

$S \leftarrow S \oplus (0^{319}\;||\;1)$

**Processing Ciphertext**

$C1 \ldots C_{t-1}\,\tilde{C}_t \leftarrow$ r-bit blocks of C, $0 \le |\tilde{C}_t| < r$

for i = 1, . . ., t − 1 do

$P_i \leftarrow S_r \oplus C_i$

$S \leftarrow C_i\;||\;S_c$

$S \leftarrow p^b(S)$

$\tilde{P}_t \leftarrow [s_r]_{|\tilde{C}_t|} \oplus \tilde{C}_t$

$S_r \leftarrow S_r \oplus (\tilde{P}_t\;||\;1\;||\;0*)$

**Finalization**

$S \leftarrow p^a\,(S \oplus (0^{r}\;||\;K\;||\;0^{320-r-k}))$

$T* \leftarrow [S]^{128} \oplus [K]^{128}$

**if** $T = T*$ return $P_1\;||\;...\;||\;P_{t-1}\;||\;P_t$

**else** return ⊥

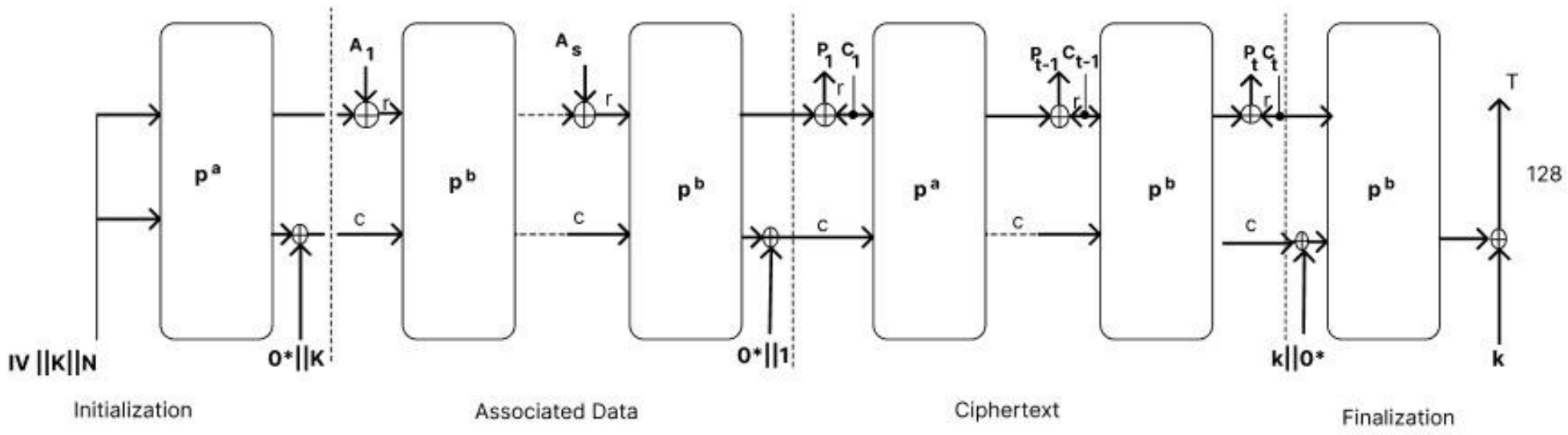

*Fig.8: Decryption in Ascon*

## 3.2. AES (Advanced Encryption Standard) encryption:

AES (Advanced Encryption Standard) is a widely used encryption algorithm renowned for its ability to secure digital data effectively. Operating on fixed-sized blocks of 128 bits arranged in a 4 × 4 column-major order array, AES employs a symmetric-key approach, where the same key is used for both encryption and decryption [Fig.9]. There are three key lengths available in AES: 128, 192, and 256 bits. Each key length corresponds to a different level of security and computational complexity.

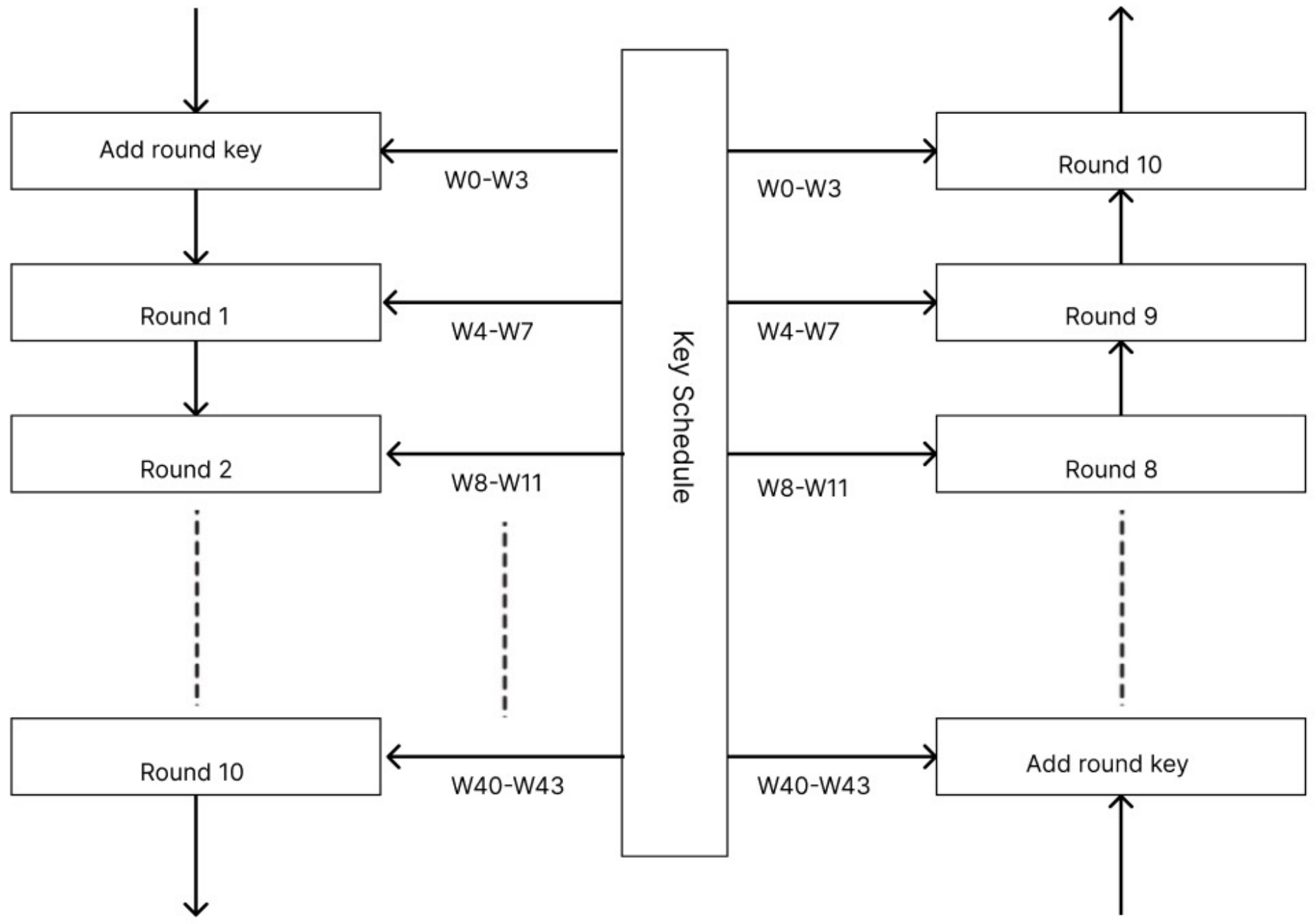

*Fig.9: AES Encryption and Decryption*

With a 128-bit key, AES-128 carries out ten encryption/decryption cycles. This version is appropriate for a wide range of applications since it finds a balance between security and performance. Although AES-128 provides enough security for the majority of use cases, because to its comparatively shorter key length, it might not be the best option for really sensitive data.

Conversely, AES-192 uses a 192-bit key and processes data in 12 cycles. Because of the longer key and more rounds than AES-128, this option offers a higher level of protection. For situations where heightened security is necessary, AES-192 is a good option because it provides more protection than AES-128 but at the expense of a minor drop in performance.

The safest version, AES-256, has a 256-bit key and goes through 14 encryption/decryption cycles. With the strongest security of the three, AES-256 is appropriate for really sensitive data since it provides strong defence against a range of cryptographic attacks. However, the computational complexity increases due to the longer keys and more rounds, which influences performance.

When efficiency is paramount, AES-128 is often chosen due to its well-balanced security-performance trade off. Whereas when a higher level of security is needed, AES-192 and AES-256 are recommended, even with the corresponding performance overhead. The exact security needs and computational limitations of the application or system ultimately determine the length of the AES key. Here, efficiency is the top priority, so AES-128 is the perfect encryption method for low-resource devices that yet need to be strong.

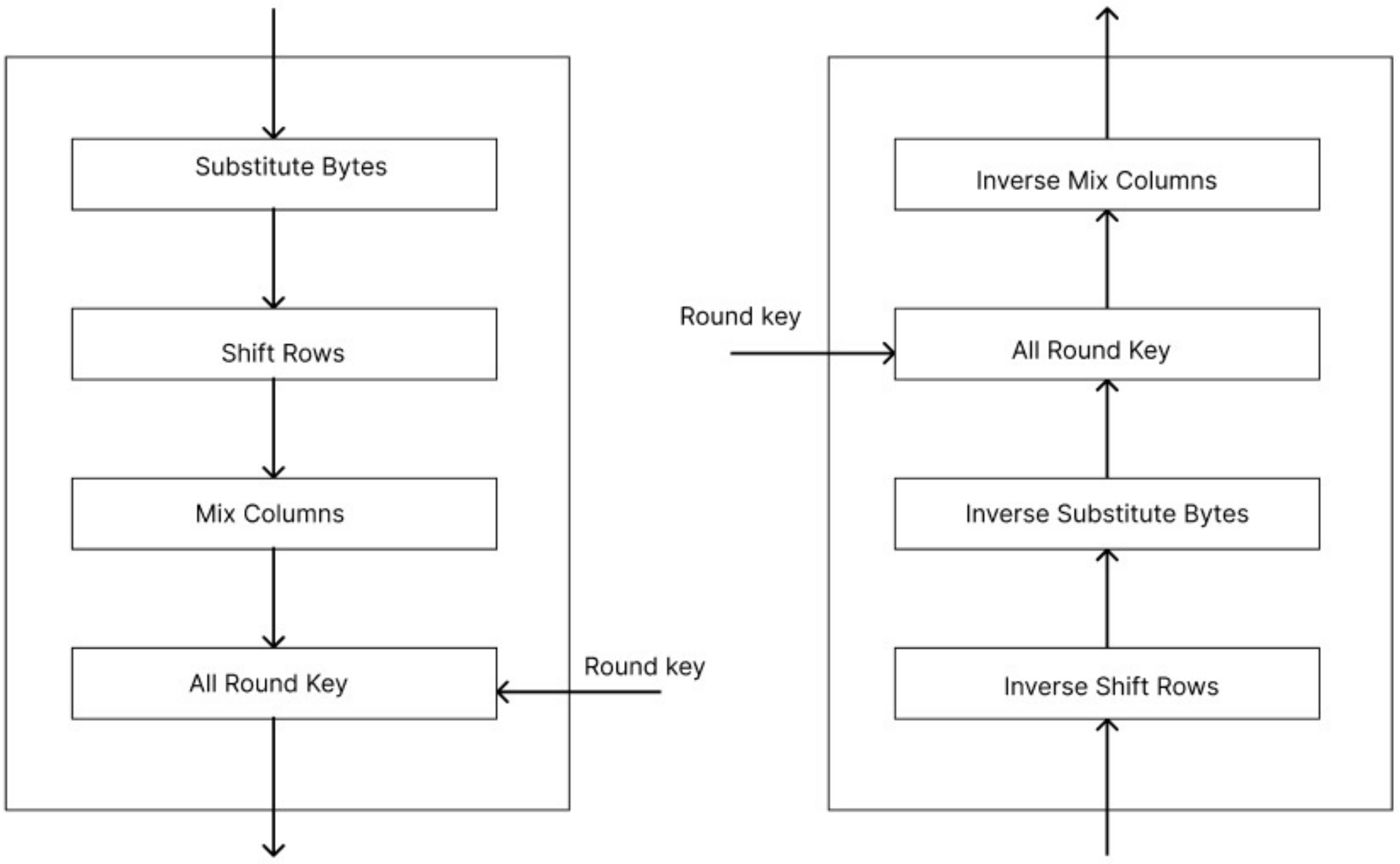

*Fig.10: Encryption Round and Decryption Round*

A. AES BYTE SUBSTITUTION (SubBytes):

Using a predefined lookup table called the S-box, the 16 input bytes are substituted in the AES Byte Substitution (SubBytes) stage. The S-box's matching value for each byte is substituted, creating a matrix with four rows and four columns. The inverse of this substitution, called InvSubBytes, is applied during decryption.

B. SHIFT ROWS NODES:

The rows of the matrix are shifted to the left using Shift Rows. The next three rows are shifted one, two, and three spaces to the left of the first row, which stays in place. The bytes are moved in relation to one another to generate a new matrix during this procedure.

C. MIX COLUMNS NODES:

With the help of a mathematical function, Mix Columns modifies each column of four bytes. Based on the original column, this function creates four entirely new bytes. Mix Columns is skipped during the last round of encryption and decryption.

D. ADD ROUND KEY NODES:

The round key, which is obtained from the encryption key, is XORed with the 16 bytes of the matrix in Add Round Key. To create the ciphertext, this phase combines the plaintext with the key. If it's the final round, the output is the ciphertext; otherwise, the process continues with another round.

E. DECRYPTION PROCESS:

As Shown in Fig. 10, During decryption the process is reversed. In reverse order, the following processes are performed in each round: Add Round Key, Mix Columns, Shift Rows, and Byte Substitution. The encryption and decryption methods are closely comparable despite the reversal, but they require different implementations because each round's subprocesses are executed in reverse order.

2.2. GOAL

The overarching goal of this research project is to advance the field of secure communication in unmanned aerial vehicle (UAV) networks through the investigation, evaluation, and implementation of lightweight cryptography algorithms, with a particular focus on ASCON. This goal is multifaceted and encompasses several interconnected objectives aimed at addressing critical challenges and advancing the state-of-the-art in UAV communication security.

A. Enhancing Security in UAV Networks: The primary objective is to enhance the security of communication within UAV networks. As UAVs become increasingly integrated into various sectors such as military operations, agriculture, infrastructure inspection, and logistics, the need for secure and resilient communication channels becomes paramount. The goal is to develop and implement cryptographic solutions that can safeguard sensitive data transmitted between UAVs, ground stations, and other networked devices from potential threats such as interception, manipulation, and unauthorized access.

B. Optimizing Resource Utilization: A key aspect of the project is to optimize resource utilization within UAV networks. Given the inherent constraints of UAV platforms, including limited computational power, memory, and energy resources, it is essential to develop cryptographic algorithms that strike a balance between security and resource efficiency. The goal is to identify lightweight cryptography solutions that can provide robust security while minimizing computational overhead, memory footprint, and energy consumption, thereby ensuring the feasibility and scalability of secure communication in UAV networks.

C. Evaluating ASCON's Suitability: A specific focus of the project is to evaluate the suitability of the ASCON encryption algorithm for securing communication in UAV networks. ASCON is a lightweight cryptographic algorithm that has shown promise in terms of both security and efficiency. The goal is to conduct a comprehensive evaluation of ASCON's performance in UAV communication scenarios, comparing it with traditional encryption algorithms such as AES. By assessing ASCON's efficacy in encrypting and decrypting communication data under various operating conditions, the project aims to determine its suitability for real-world UAV applications.

D. Contributing to Research and Development: Another goal of the project is to contribute to the broader research and development efforts in the field of UAV communication security. By conducting rigorous experiments, analyzing performance metrics, and synthesizing research findings, the project aims to generate new insights, methodologies, and best practices for designing secure and efficient communication protocols tailored to the unique requirements of UAV platforms. The goal is to provide actionable recommendations and guidelines that can inform future research directions and industry practices, ultimately advancing the state-of-the-art in UAV communication security.

E. Facilitating Adoption of Lightweight Cryptography: Ultimately, the overarching goal of the project is to facilitate the adoption of lightweight cryptography algorithms, such as ASCON, in UAV communication systems. By demonstrating the effectiveness, efficiency, and security of ASCON in simulated UAV network environments, the project aims to provide empirical evidence and practical guidance for integrating lightweight cryptography solutions into real-world UAV applications. The goal is to empower UAV operators, developers, and researchers to make informed decisions regarding cryptographic algorithm selection and implementation, thereby promoting the widespread adoption of secure and resource-efficient communication frameworks in the UAV technology domain.

## 4. DESIGN APPROACH AND DETAILS

Our approach encompasses a systematic investigation into the performance and efficiency of cryptographic algorithms, specifically focusing on Ascon and AES-128. We begin by implementing these algorithms in Python, meticulously adhering to their respective specifications. This ensures accuracy and reliability in our evaluations. Leveraging the functionality of the `Crypto` library for AES-128 and developing Ascon variants from scratch, we create a robust framework capable of encryption, decryption, hashing, and message authentication code (MAC) operations. To evaluate performance, we conduct extensive experimentation, measuring execution time and throughput across varying input data sizes. Our experimental setup includes a diverse dataset of random plaintext values stored in an Excel file, ensuring comprehensive testing. Through multiple iterations and rigorous analysis, we derive insights into the comparative efficiency of each algorithm variant. Visualization techniques aid in presenting our findings, facilitating clear comparisons and conclusions. Furthermore, we validate our implementation against established cryptographic standards to ensure correctness and reliability. Despite inherent limitations, such as computational resource constraints, our approach lays a solid foundation for future optimizations and real-world application considerations. Through this approach, we aim to contribute valuable insights to the field of cryptography, fostering advancements in security and privacy across diverse computing environments.

### 4.1. CONSTRAINTS, ALTERNATIVES AND TRADE-OFFS

In the pursuit of evaluating cryptographic algorithms, we encounter various constraints that shape our approach and decision-making process. One significant constraint is the limitation of computational resources, including CPU power and memory availability. Due to the unavailability of abundant resources, we are unable to perform in-depth security analysis, which necessitates a cautious approach to algorithm selection and implementation. To address the constraints posed by limited computational resources, we explore alternative strategies such as algorithmic optimizations and hardware acceleration. Algorithmic optimizations involve refining implementation techniques to minimize computational overhead, enhance throughput, and reduce memory footprint. By streamlining algorithms and reducing unnecessary computations, we aim to improve overall efficiency without requiring additional hardware resources.

Additionally, we consider hardware acceleration as an alternative approach to mitigate the impact of resource constraints. This involves leveraging specialized cryptographic hardware or utilizing GPUs to offload intensive computations to dedicated processing units. By harnessing the parallel processing capabilities of hardware accelerators, we can achieve significant performance gains while minimizing the burden on CPU resources. However, embracing these alternatives introduces trade-offs that must be carefully considered. Algorithmic optimizations, while improving efficiency, may lead to code complexity and reduced readability, making future modifications and debugging more challenging. Similarly, hardware acceleration solutions require upfront investment in specialized hardware or infrastructure, which may not be feasible for all deployment scenarios.

Moreover, trade-offs extend beyond technical considerations to encompass security implications and cryptographic properties. While algorithmic optimizations and hardware acceleration can enhance performance, they may inadvertently introduce vulnerabilities or weaken cryptographic guarantees if not implemented carefully. Balancing the pursuit of efficiency with the need for robust security is essential to maintain the integrity and confidentiality of encrypted data.

In navigating these constraints, alternatives, and trade-offs, our approach emphasizes a comprehensive evaluation that considers not only computational efficiency but also security, usability, and practicality. By carefully weighing the pros and cons of different strategies, we aim to develop a balanced and resilient cryptographic framework that meets the diverse requirements of modern computing environments while upholding the highest standards of security and reliability.

## 5. RESULT ANALYSIS

Performance metrics for various cryptographic algorithms are collected continuously throughout communication. It measures the amount of time required for each cryptographic operation using timers or performance counters. It computes total throughput, communication delay, and average and peak processing times. The module tracks the latency encountered in UAV communication, including sending, processing, and receiving times, and analyses throughput both overall and per execution.

A. PEAK TIME: represents the maximum execution time observed among all executions for a specific process, is calculated by finding the maximum among the execution times measured for each run of the process. Where $t_1, t_2, \ldots, t_n$ are the execution times measured for each run of the process.

$$\text{Peak Time} = \text{Max}\ (t_1, t_2, \ldots, t_n)$$

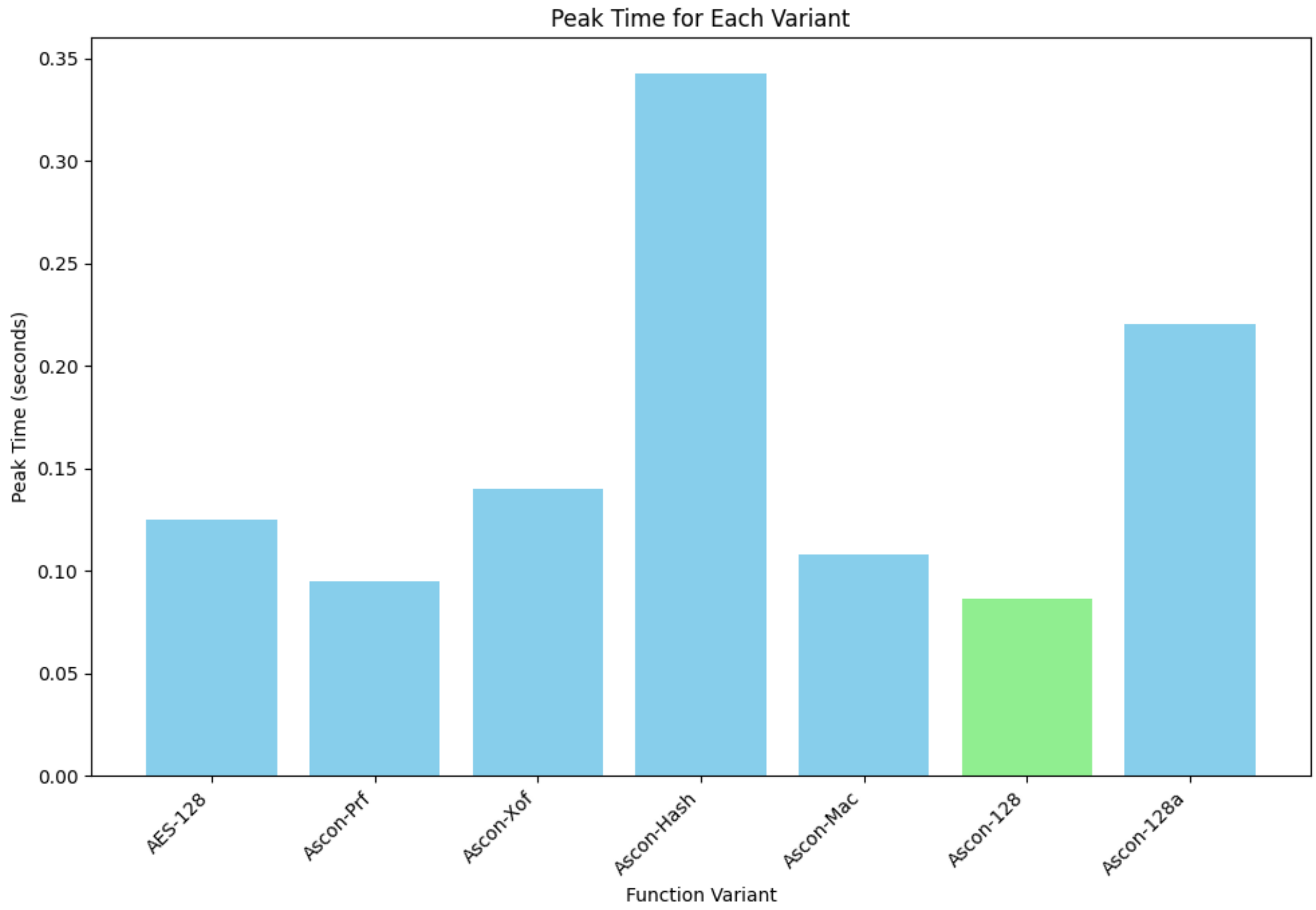

*Fig.11: Peak Time for Each Algorithm*

The peak execution time for each cryptographic algorithm is listed in a Fig.11. The Fig.11 shows that Ascon-Hash has the highest peak execution time (0.342991 seconds), while Ascon-128a has the second highest peak execution time (0.220283 seconds).

B. <u>AVERAGE TIME:</u> Average Time, on the other hand, signifies the average execution time observed among all executions for a particular process. It is determined by summing up all the execution times and dividing by the total number of executions. Where $t_1, t_2, \ldots, t_n$ are the execution times measured for each run of the process, and n is the total number of executions.

$$\text{Average Time} = (t_1, t_2, \ldots, t_n) / n$$

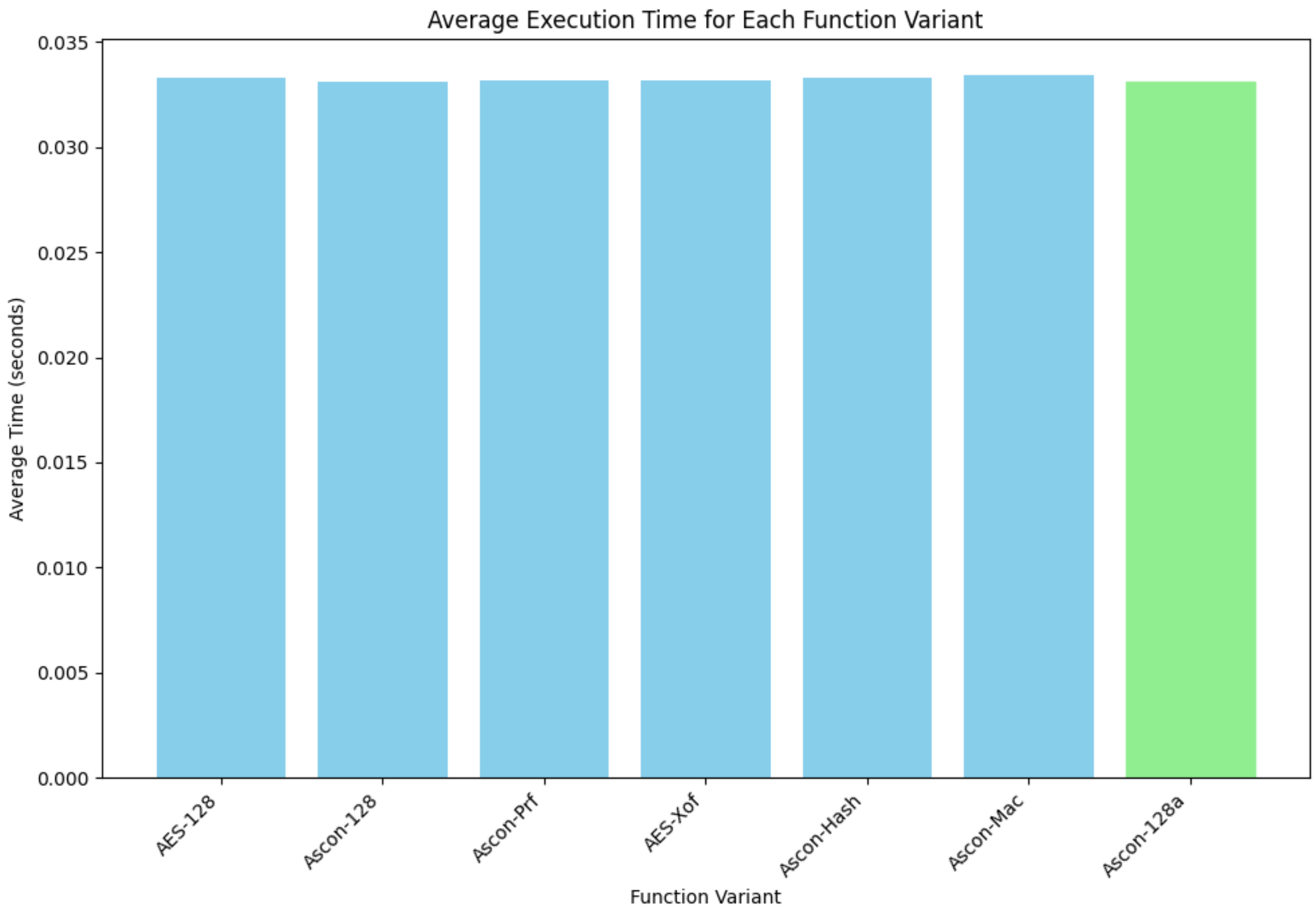

*Fig.12: Average Execution Time for Each Algorithm/Variant in Bar Graph*

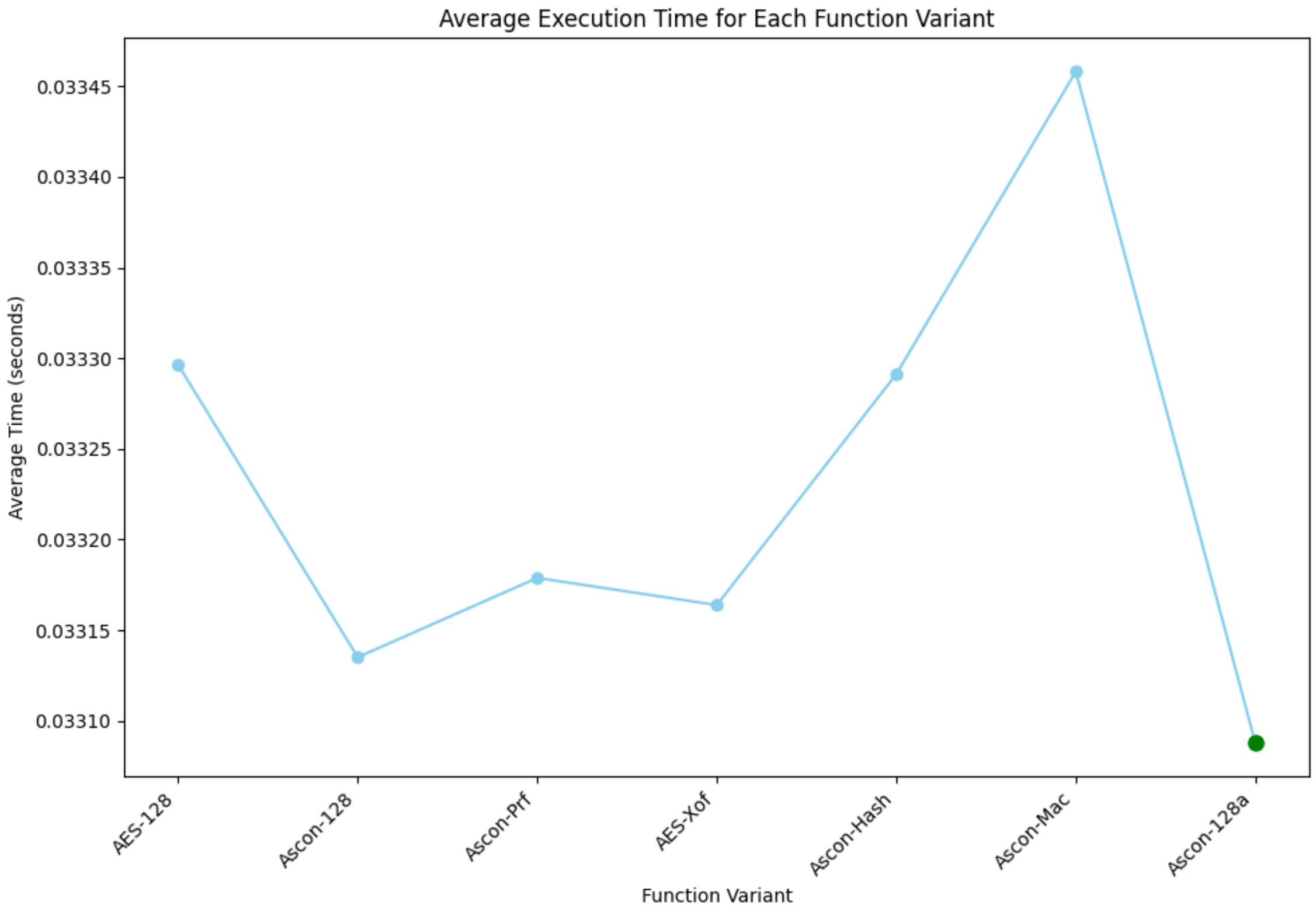

*Fig.13: Average Execution Time for Each Algorithm/Variants in Line Graph*

The average execution time [Fig.12 and Fig.13] for each cryptographic algorithm is presented in a line graph and a bar graph (as shown in the image). The x-axis represents the cryptographic algorithm, and the y-axis represents the average execution time in seconds. The data shows that Ascon-128a has the lowest average execution time (0.033088 seconds), while Ascon-Hash has the highest average execution time (0.033164 seconds).

C. <u>OVERALL THROUGHPUT:</u> It is crucial for assessing the rate at which data is processed by a system over a given time period, is typically measured in bytes per second (B/s). This metric is calculated by dividing the total amount of data processed by the total processing time for that specific process. Where Total Data Processed is sum of the sizes of all input data processed by the process and Total Processing Time is sum of the execution times for all executions of the process.

| Overall Throughput=Total Processing Time of all inputs/Total Data Processed |
| --- |

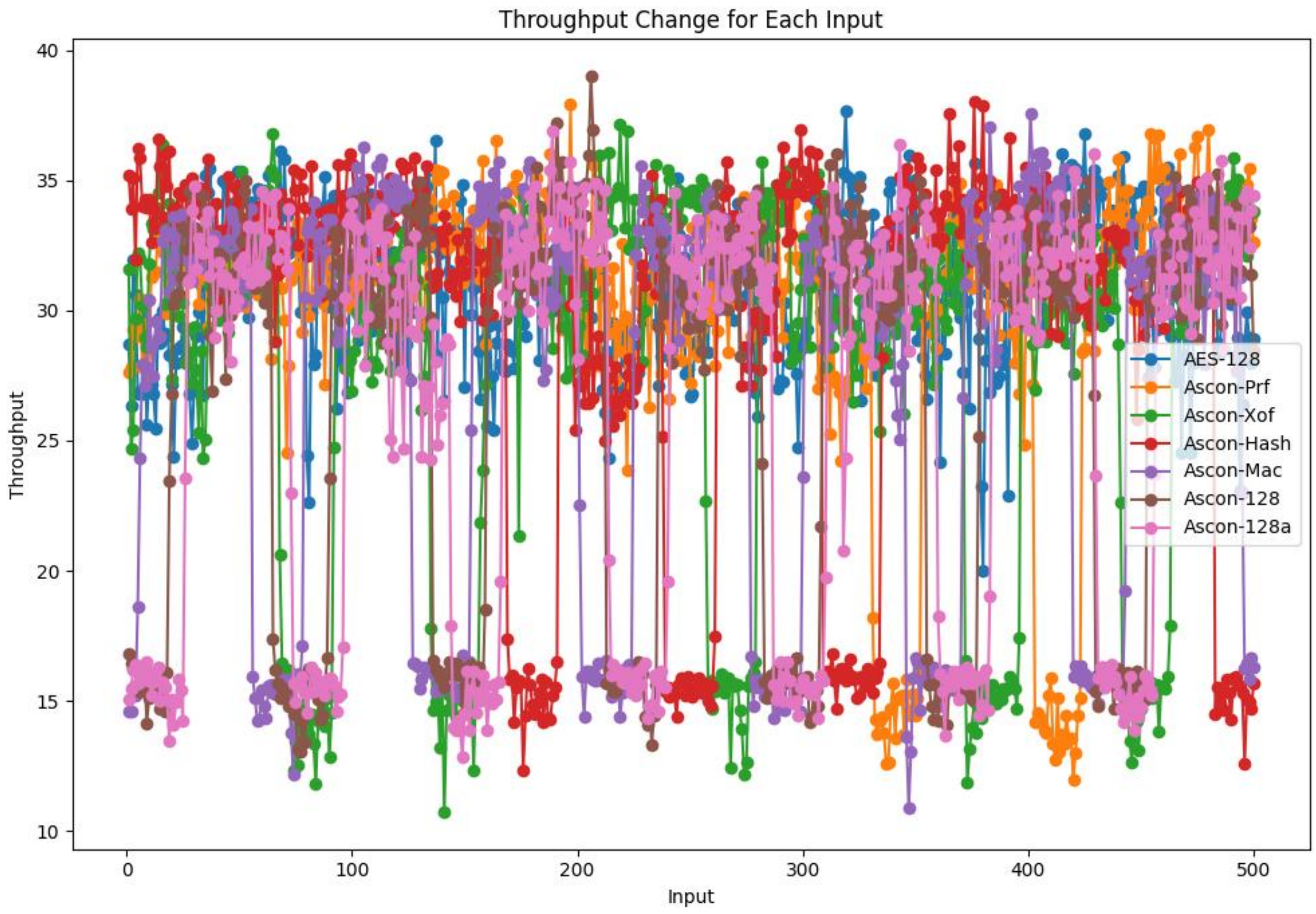

*Fig.14: Throughput change in Each Algorithm*

This fig.14 shown the change in throughput change in the whole process of passing all input of dataset through encryption and Decryption process for each Algorithm.

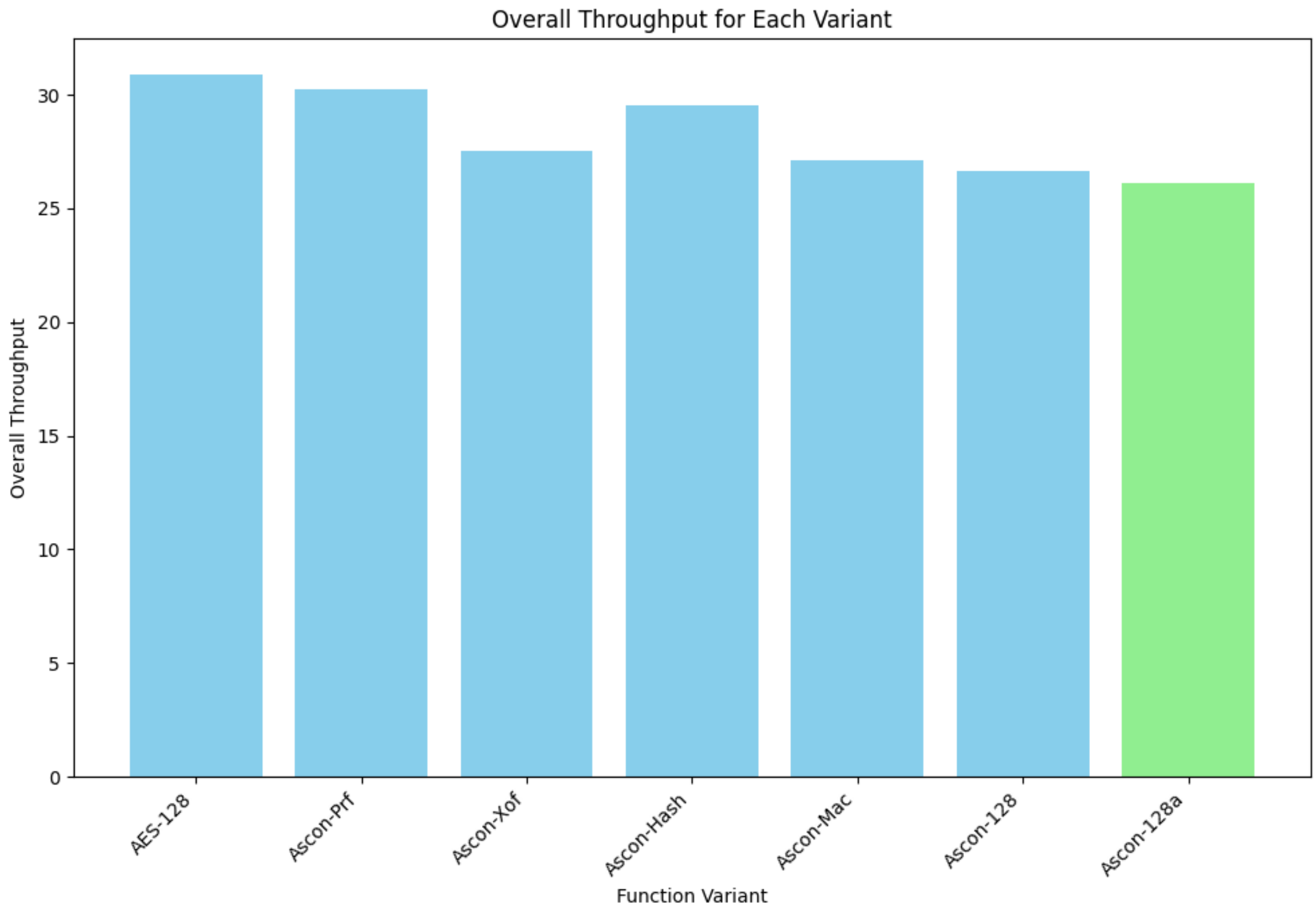

*Fig.15: Overall Throughput for Each Algorithm*

Here, we attempt to determine the follow-up of throughput change for the entire dataset as well as the overall throughput for each algorithm, which is an average of throughput change. For the most part, most algorithms have extremely comparable execution times. The average execution time of all algorithms is 0.033 seconds [Fig.15]. This implies that, for machines with limited resources, they are all fairly efficient. With an average execution time of 0.033088 seconds, Ascon-128a has the lowest. This suggests that, out of all the examined algorithms, it may be the fastest processing choice. With an average execution time of 0.033164 seconds, Ascon-Hash has the longest. Ascon-Hash might be marginally less efficient than other algorithms, despite the modest difference.

## 5.1. SECURITY ANALYSIS

When assessing the resistance of Ascon-128a and AES-128 to different cryptographic attacks, each cipher presents unique strengths and vulnerabilities that affect their suitability for securing UAV communications.

Ascon-128a stands out for its efficiency and robust cryptographic design, making it particularly resistant to implementation-based attacks like fault analysis [38] [39] [42] and side-channel attacks [40]. Its tailored cryptographic features and streamlined architecture enhance its resilience against these attacks, ensuring the confidentiality and integrity of data transmissions in resource-constrained environments such as UAV networks. Additionally, Ascon-128a's performance metrics demonstrate its effectiveness in real-time communication scenarios, further bolstering its suitability for UAV communication security. On the other hand, AES-128 has undergone extensive refinement and adaptation to address classical cryptanalytic techniques such as differential and linear cryptanalysis. With ongoing research efforts throughout all these years, AES-128 has been strengthened to withstand these attacks, with updates to its key expansion algorithm and other components. Its widespread adoption and standardized implementation make it a trusted choice in security-sensitive applications, providing a strong defence against known cryptanalytic methods.

| **Cryptographic Attack** | **AES-128** | **ASCON-128a** | **references** |
|---|---|---|---|
| fault analysis | ✓ (not much susceptible) | ✓ | [38] [39] [42] [43] |
| side-channel attacks | | ✓ | [40] |
| differential cryptanalysis | ✓ | | [41] |
| linear cryptanalysis | ✓ | | [41] |

*Table 3: AES-128 and ASCON-128a resistance to Cryptographic Attack*

In conclusion [Table3], while Ascon-128a excels in resilience against implementation-based attacks [41] like fault analysis, AES-128 demonstrates robustness against classical cryptanalytic techniques. The choice between the two ciphers for UAV communication security depends on the specific threat landscape and operational requirements, ensuring a balanced approach to achieving security, efficiency, and practicality in UAV communication systems.

## 5.2. COMPARATIVE STUDY

After conducting an in-depth analysis and considering various factors, including performance metrics, security properties, and suitability for resource-constrained environments, the research results indicate that Ascon-128a emerges as the optimal choice for UAV communication systems requiring a balance between efficiency and security.

A. In terms of Performance Metrics: Ascon-128a demonstrates superior performance metrics compared to AES-128, particularly in terms of average execution time. The low average execution time of Ascon-128a highlights its efficiency in processing cryptographic operations, making it well-suited for resource-constrained environments where processing speed is crucial. Despite its slightly higher peak execution time compared to AES-128, Ascon-128a's overall performance efficiency remains unmatched.

B. In terms of Security Properties: While AES-128 boasts broadly resilience to attacks and continuous refinement, Ascon-128a offers robust security features tailored for real-time communication scenarios. Ascon-128a provides authenticated encryption capabilities, ensuring both confidentiality and integrity of data transmissions. Its streamlined design and cryptographic strength make it a compelling option for securing UAV communication systems, especially in scenarios where rapid cryptographic processing is essential.

C. Suitability for Resource-Constrained Environments: In resource-constrained environments, where computational resources are limited, the efficiency of cryptographic algorithms becomes most significant. Ascon-128a's efficient processing capabilities, coupled with its strong security profile, position it as the preferred choice for UAV communication systems operating under resource constraints. By offering a balance between speed and security, Ascon-128a addresses the specific needs of resource-constrained environments, ensuring optimal performance without compromising on security assurance.

## 6. CONCLUSION

After thorough consideration of all factors, including performance, security, and suitability for resource-constrained environments, Ascon-128a emerges as the superior choice for UAV communication systems. While AES-128 offers resilience to attacks and continuous refinement, Ascon-128a's efficiency and tailored cryptographic features make it the optimal solution for scenarios requiring fast and secure cryptographic operations in resource-constrained environments. By leveraging the strengths of Ascon-128a, UAV communication systems can achieve optimal performance and security, ensuring reliable and secure communication in challenging operational environments.

## Conflict of Interest

The authors declare that they have no conflicts of interest related to this research.

## Funding Information

No funding is received to conduct this work.

## Author contribution

Aanchal Patel has conducted the study, experimental analysis and prepared the draft.

Aswani Kumar Cherukuri has supervised the work, verified the results, corrected the draft.

## Data Availability Statement

Not applicable

## Research Involving Human and /or Animals Informed Consent

Not applicable

## REFERENCES:

1. "NIST selects 'Lightweight Cryptography' algorithms to protect small devices | NIST," NIST, Feb. 07, 2023. https://www.nist.gov/news-events/news/2023/02/nist-selects-lightweight-cryptography-algorithms-protect-small-devices
2. N. Cecchinato, A. Toma, C. Drioli, G. Oliva, G. Sechi and G. L. Foresti, "Secure Real-Time Multimedia Data Transmission from Low-Cost UAVs with A Lightweight AES Encryption," in IEEE Communications Magazine, vol. 61, no. 5, pp. 160-165, May 2023, doi: 10.1109/MCOM.001.2200611. keywords: {Wireless communication; Military computing; Data security; Streaming media; Real-time systems; Microphone arrays; Encryption},
3. Computer Security Division, Information Technology Laboratory, National Institute of Standards and Technology, U.S. Department of Commerce, "Finalists - Lightweight Cryptography | CSRC | CSRC." https://csrc.nist.gov/Projects/lightweight-cryptography/finalists
4. L. Gupta, R. Jain and G. Vaszkun, "Survey of Important Issues in UAV Communication Networks," in IEEE Communications Surveys & Tutorials, vol. 18, no. 2, pp. 1123-1152, Secondquarter 2016, doi: 10.1109/COMST.2015.2495297. keywords: {Mobile computing; Topology; Network topology; Mesh networks; Routing protocols; Routing; Unmanned Aerial Vehicle;UAV; Multi-UAV Networks;

ad hoc networks; communication networks; wireless mesh networks; software defined network; routing; seamless handover; energy efficiency},

5. T. Li et al., "Lightweight Secure Communication Mechanism Towards UAV Networks," 2019 IEEE Globecom Workshops (GC Wkshps), Waikoloa, HI, USA, 2019, pp. 1-6, doi: 10.1109/GCWkshps45667.2019.9024530. keywords: {Encryption; Drones; Packet loss; Authentication},
6. Kamil, Muhammed Zahid. NOVEL LIGHTWEIGHT ENCRYPTION API FOR IOT DEVICE COMMUNICATION. Diss. California State Polytechnic University, Pomona, 2023.
7. Abdullah, Ako Muhamad. "Advanced encryption standard (AES) algorithm to encrypt and decrypt data." Cryptography and Network Security 16.1 (2017): 11.
8. N. Cecchinato, A. Toma, C. Drioli, G. Oliva, G. Sechi and G. L. Foresti, "Secure Real-Time Multimedia Data Transmission from Low-Cost UAVs with A Lightweight AES Encryption," in IEEE Communications Magazine, vol. 61, no. 5, pp. 160-165, May 2023, doi: 10.1109/MCOM.001.2200611. keywords: {Wireless communication; Military computing; Data security; Streaming media; Real-time systems; Microphone arrays; Encryption},
9. Cecchinato N, Toma A, Drioli C, Oliva G, Sechi G, Foresti GL. A Secure Real-time Multimedia Streaming through Robust and Lightweight AES Encryption in UAV Networks for Operational Scenarios in Military Domain. Procedia Computer Science. 2022;205:50-57. doi:10.1016/j.procs.2022.09.006
10. Zhong Y, Gu J. Lightweight block ciphers for resource-constrained environments: A comprehensive survey. Future Generation Computer Systems. 2024;157:288-302. doi:10.1016/j.future.2024.03.054
11. Turan, Meltem Sonmez, et al. Status report on the final round of the NIST lightweight cryptography standardization process. US Department of Commerce, National Institute of Standards and Technology, 2023.
12. Turan, Meltem Sönmez, et al. "Status report on the first round of the NIST lightweight cryptography standardization process." National Institute of Standards and Technology, Gaithersburg, MD, NIST Interagency/Internal Rep.(NISTIR) 108 (2019).
13. Turan, Meltem Sonmez, et al. "Status report on the second round of the NIST lightweight cryptography standardization process." (2021).

14. McKay, Kerry, et al. Report on lightweight cryptography. No. NIST Internal or Interagency Report (NISTIR) 8114 (Draft). National Institute of Standards and Technology, 2016.
15. Mohajerani, Kamyar, et al. "Sca evaluation and benchmarking of finalists in the nist lightweight cryptography standardization process." Cryptology ePrint Archive (2023).
16. I. Elsadek, S. Aftabjahani, D. Gardner, E. MacLean, J. R. Wallrabenstein and E. Y. Tawfik, "Hardware and Energy Efficiency Evaluation of NIST Lightweight Cryptography Standardization Finalists," 2022 IEEE International Symposium on Circuits and Systems (ISCAS), Austin, TX, USA, 2022, pp. 133-137, doi: 10.1109/ISCAS48785.2022.9937643
17. El-Hajj, Mohammed, Hussien Mousawi, and Ahmad Fadlallah. "Analysis of Lightweight Cryptographic Algorithms on IoT Hardware Platform." Future Internet 15.2 (2023): 54
18. Mattsson, John Preuß, et al. "Proposals for Standardization of the Ascon Family."
19. Dobraunig, Christoph, et al. "Ascon PRF, MAC, and short-input MAC." Cryptology ePrint Archive (2021)
20. M. Tanveer, A. U. Khan, H. Shah, S. A. Chaudhry and A. Naushad, "PASKE-IoD: Privacy-Protecting Authenticated Key Establishment for Internet of Drones," in IEEE Access, vol. 9, pp. 145683-145698, 2021, doi: 10.1109/ACCESS.2021.3123142.
21. Chakraborty, Bishwajit, Chandranan Dhar, and Mridul Nandi. "Exact Security Analysis of ASCON." Cryptology ePrint Archive (2023)
22. Kamil, Muhammed Zahid. NOVEL LIGHTWEIGHT ENCRYPTION API FOR IOT DEVICE COMMUNICATION. Diss. California State Polytechnic University, Pomona, 2023.
23. J. Sliwa, K. Wrona, T. Shabanska and A. Solmaz, "Lightweight quantum-safe cryptography in underwater scenarios," 2023 IEEE 48th Conference on Local Computer Networks (LCN), Daytona Beach, FL, USA, 2023, pp. 1-6, doi: 10.1109/LCN58197.2023.10223321
24. Gebremariam, T. H. Digital Twin and Securing IoT Applications in Industry 4.0. MS thesis. University of Twente, 2023
25. V. K. Ralegankar et al., "Quantum Cryptography-as-a-Service for Secure UAV Communication: Applications, Challenges, and Case Study," in IEEE Access, vol. 10, pp. 1475-1492, 2022, doi: 10.1109/ACCESS.2021.3138753. keywords: {Autonomous aerial vehicles; Security; Blockchains; Cryptography; Drones; Military communication; Medical services; Unmanned aerial vehicle; quantum computing; quantum cryptography; military; blockchain},

26. Sharma J, Mehra PS. Secure communication in IOT-based UAV networks: A systematic survey. Internet of Things. 2023;23:100883. doi:10.1016/j.iot.2023.100883
27. Mekdad, Yassine, et al. "A survey on security and privacy issues of UAVs." Computer Networks 224 (2023): 109626
28. Xu, Fang, et al. "Beyond encryption: Exploring the potential of physical layer security in UAV networks." Journal of King Saud University-Computer and Information Sciences (2023): 101717
29. T. Alladi, Naren, G. Bansal, V. Chamola and M. Guizani, "SecAuthUAV: A Novel Authentication Scheme for UAV-Ground Station and UAV-UAV Communication," in IEEE Transactions on Vehicular Technology, vol. 69, no. 12, pp. 15068-15077, Dec. 2020, doi: 10.1109/TVT.2020.3033060. keywords: {Authentication; Protocols; Unmanned aerial vehicles; Computational modeling; Physical unclonable function; Cryptography; Internet of Drones (IoD); PUFs; UAVs; mutual authentication; physical security; privacy; security protocol},
30. Mohd Noor, Norzailawati, Alias Abdullah, and Mazlan Hashim. "Remote sensing UAV/drones and its applications for urban areas: A review." IOP conference series: Earth and environmental science. Vol. 169. IOP Publishing, 2018.
31. O. Sami Oubbati, M. Atiquzzaman, T. Ahamed Ahanger and A. Ibrahim, "Softwarization of UAV Networks: A Survey of Applications and Future Trends," in IEEE Access, vol. 8, pp. 98073-98125, 2020, doi: 10.1109/ACCESS.2020.2994494. keywords: {Unmanned aerial vehicles; 5G mobile communication; Security; Wireless communication; Routing protocols; Routing; Cellular networks; UAV; SDN; NFV; 5G; B5G; Cellular networks},
32. Schläffer CDME Florian Mendel, Martin. ASCON – Authenticated Encryption and Hashing. https://ascon.iaik.tugraz.at/index.html.
33. Dobraunig C, Eichlseder M, Mendel F, Schläffer M. ASCON v1.2: Lightweight Authenticated Encryption and Hashing. Journal of Cryptology. 2021;34(3). doi:10.1007/s00145-021-09398-9
34. Schläffer M, Ascon. *Ascon*.; 2021. https://ascon.iaik.tugraz.at/files/asconv12-nist.pdf.
35. Dobraunig C, Eichlseder M, Mendel F, et al. Ascon V1.2. In: Submission to the CAESAR Competition.; 2016. https://ascon.iaik.tugraz.at/files/asconv12.pdf.

36. Sharma A, Vanjani P, Paliwal N, et al. Communication and networking technologies for UAVs: A survey. Journal of Network and Computer Applications. 2020;168:102739. doi:10.1016/j.jnca.2020.102739
37. Mekdad Y, Arış A, Babun L, et al. A survey on security and privacy issues of UAVs. Computer Networks. 2023;224:109626. doi:10.1016/j.comnet.2023.109626
38. K. Ramezanpour, P. Ampadu and W. Diehl, "A Statistical Fault Analysis Methodology for the Ascon Authenticated Cipher," 2019 IEEE International Symposium on Hardware Oriented Security and Trust (HOST), McLean, VA, USA, 2019, pp. 41-50, doi: 10.1109/HST.2019.8741029. keywords: {Ciphers; Encryption; Sparse matrices; Hamming weight; Authentication; Authenticated encryption; Ascon; CAESAR; fault injection; ineffective fault; SIFA; statistical fault analysis},
39. Joshi P, Mazumdar B. SSFA: Subset fault analysis of ASCON-128 authenticated cipher. Microelectronics Reliability/Microelectronics and Reliability. 2021;123:114155. doi:10.1016/j.microrel.2021.114155
40. Luo S, Wu W, Li Y, Zhang R, Liu Z. An efficient soft analytical Side-Channel attack on ASCon. In: Lecture Notes in Computer Science. ; 2022:389-400. doi:10.1007/978-3-031-19208-1_32
41. View of Bounds for the Security of Ascon against Differential and Linear Cryptanalysis. https://tosc.iacr.org/index.php/ToSC/article/view/9527/9064.
42. Mestiri H, Barraj I, Bedoui M, Machhout M. An ASCON AOP-SystemC environment for security fault analysis. Symmetry. 2024;16(3):348. doi:10.3390/sym16030348
43. Floissac N, L'Hyver Y, SERMA TECHNOLOGIES ITSEF. From AES-128 to AES-192 and AES-256, How to Adapt Differential Fault Analysis Attacks.; 2021. https://eprint.iacr.org/2010/396.pdf.